# An outlook on the Rapid Decline of Carbon Sequestration in French Forests and associated reporting needs

P. Ciais[1], C. Zhou[1], P. Schneider[2,3,1], M. Schwartz[1], N. Besic[4], C. Vega[4] and J.-D. Bontemps[4]

1 Laboratoire des Sciences du Climat et de l'Environnement, LSCE/IPSL, CEA-CNRS-UVSQ, Université Paris Saclay, 91191 Gif-sur-Yvette, France

2 Swiss Federal Institute for Forest, Snow and Landscape Research (WSL), Forest Dynamics, Birmensdorf, Switzerland

3 ETH Zurich (Swiss Federal Institute of Technology), Institute of Terrestrial Ecosystems, Zurich, Switzerland

4 IGN, ENSG, Laboratoire d'Inventaire Forestier (LIF), 54000 Nancy, France

**Abstract**

In this study, we present and discuss changes in carbon storage in the French forests from 1990 to 2022, derived from CITEPA statistics on forest carbon accounting, fed by National Forest Inventory (NFI) data collected through an extensive network of measurement sites across Metropolitan France, and other data sources as regards forest removals. The NFI is designed to provide statistical estimations of growing stock, gains and losses at the national or subnational levels but is unable, in its classical form, to provide detailed spatial outlook, such as on abrupt losses during fires, droughts and insect attacks. A continuing removal of $CO_2$ from the atmosphere by the French forests occurred from 1990 to 2022, because harvest and mortality $CO_2$ losses remained smaller than $CO_2$ removals by forest growth and the increase in forest area (70,000 ha per year but insignificant in terms of increased carbon stocks at present). The $CO_2$ removal by forests was 49.3 $MtCO_2$ $yr^{-1}$ in 1990, increased to reach a peak of 74.1 $MtCO_2$ $yr^{-1}$ in 2008 and then quickly decreased down to 37.8 Mton $CO_2$ $yr^{-1}$ in 2022. The changes of $CO_2$ removal by forests can be separated in three phases. From 1990 to 2013, the $CO_2$ removal increased alongside increasing growth of living trees. A spike of carbon loss was caused by the passage of the Lothar and Martin extra-tropical cyclones but forests recovered rapidly within a few years. In contrast, from 2013 to 2017, the $CO_2$ removal by forests quickly decreased quickly due to increasing $CO_2$ losses from harvest and natural mortality, each process contributing almost equally. After 2017, the sink remained low and mortality rates stayed larger than during any of the previous years. This recent period is marked by climate shocks such as summer droughts and heatwaves in 2015, 2018, 2022, 2023. The full impacts of the droughts in 2022 and 2023 are not yet covered with full precision, as some of the sites measured by the national inventory before those droughts are still pending a second visit. Further, lagged mortality can cause an increase of tree mortality some years after a drought happens. The different regions of France show contrasted trajectories. Southern Mediterranean regions where forests have a low harvest rate have experienced a lower increase in mortality and a sustained $CO_2$ uptake. Despite high harvest intensities, the Landes plantations also show an increasing $CO_2$ sink. In contrast, all northern regions and Corsica have sawn a strong decline in their $CO_2$ removal rates, except in the Ile-de-France region (larger Paris area), where the $CO_2$ sink was constant during the last 30 years, possibly because many forests are used for recreation and are subjected to low harvest pressure. Two regions, the Hauts de France and Grand Est forests, stand out as becoming net emitters of $CO_2$ to the atmosphere. Other regions where the $CO_2$ sink declined and is now close to zero are Normandy, Corsica, and Bourgogne-Franche Comté. We present a detailed analysis of where trees are dying in France, where mortality has increased, and which species and tree sizes are affected the most. Remarks on how sample based statistical estimation of forest carbon changes using the classical NFI approach, limited to larger administrative scales, can be complemented by new high resolution information from satellites and denser monitoring of mortality processes is presented in the end. In view of the limited estimation accuracy of some forest fluxes over the long term, we last draw recommendations to strengthen forest C sink quantification in the near future.

**Introduction**

Forests in France have been absorbing carbon over at least the past 150 years, mainly due to favorable demography with relatively young stands, following the plantations of new forests and the expansion of trees over previous agricultural lands, in particular in mountain regions and southern France (Lambin and Meyfroidt, 2011) (Denardou-Tisserand, 2019). Early afforestation for land protection started in the 19th century (Landes, Sologne, mountain restoration) and the long-lasting policy of coppice conversion into standard forests, initiated in the 1830, has led to increased forest density. During this period, harvest pressure has lowered due to fossil fuel utilization, and remained relatively modest. Natural mortality rates were low. Notably, there was a shift from coppice forestry to silviculture focused on quality timber production. This transition led to a significantly larger average stock per hectare, as the demand for fuelwood declined with the development of alternative energy sources. Consequently, as wood removal from harvest and wood losses from natural mortality remained lower than the growth of forests, the result was an increased carbon storage (Bontemps et al., 2020), also footprinted in statistics of European forests (Bontemps, 2021).

The French national forest inventory aims to provide a reliable estimation of wood volume and their changes through time by sampling trees at systematic locations. Since it does not yet produce official forest carbon statistics, these estimations are therefore translated into biomass and carbon by CITEPA (https://www.citepa.org/) for national and international reporting purposes, using prior expertises on the issue (e. g. CARBOFOR project, Loustau 2004;http://www.gip-ecofor.org/gicc/wp-content/uploads/2021/12/Projet3_rapport_final_Lousteau.pdf). Also, the NFI sampling design has changed over time. Between 1990 and 2004, only decennial, departmental, and asynchronous inventories existed. An extensive systematic annual network of non-permanent sites was established 2005 to meet the standard of annual forest inventory and strengthen the forest monitoring frequency (Bontemps and Bouriaud, 2024). Since 2010, the same sites have been revisited once after five years (notion of semi-permanent plot) to better estimate forest fluxes, while systematic sites are being renewed and measured each year for future sampling. Consequently, there is no homogeneity of the approach over the time period covered by this study.

A sampling location is also called a 'forest plot' and has a radius of 15 meters (0.1 ha) and typically contains between 5 and 100 trees. During the first visit, at each location, measurements are taken of the diameters of the censusable trees at a conventional breast height (7.5 cm diameter), while the height is measured only for the subsample of the latter, with an imputation of the tree heights that are not measured. Using allometric equations that allow calculation of the volume of each tree from its height and diameter, the volume of standing wood in the sampled forest is obtained. Growth is estimated using tree coring at breast height, focused on the last five years of growth, and is combined with volume allometries to estimate volume growth. Importantly, the information about the growth does not come from the difference in the circumference between the N+5 and N visit, though it should be the case in the future. For the moment, it is derived from tree coring at the N visit (last 5 years growth), supplemented by the total volume of new recruitments at the N+5 visit (trees that have grown above a threshold diameter in the meanwhile). The trees which were established as cut, dead or windthrown (at the N+5 visit) also make part of the "production" variable, with the assumption that they were still growing during 2.5 years before being cut. The inventory field measurements make a census of the trees that died from natural causes, i.e., were alive during the first visit but were found dead during the second one. Of note, dead trees that are harvested before the second visit (as can be the case in e. g. massive decline events such as bark beetle attacks) will not contribute to the mortality, instead to the harvest flux. As such, massive mortality can therefore drive harvest increases without being identified as such. The census reports the percentage of trees that were harvested with evidence of being cut and removed between the first and the second visit. Since 2010, each year, up to 7000 plots are newly sampled based on a predefined sampling plan, and up to 7000 plots are re-visited from the same locations that were sampled for the first time 5 years before, thus in total, ca. 14000 plots are measured each year.

Of importance, harvest measurements before 2010 have long been considered unreliable by the NFI, as the stump inventory and dating protocol on temporary plots was inaccurate enough. Recent research works (Audinot, 2021; Denardou-Tisserand, 2019) have confirmed their biasedness, and historical underestimation by 50%. For those reasons, harvest rates have been assessed from external data sources by CITEPA before 2010, including the *Enquête Annuelle de Branche* from the Agricultural statistics based on wood fluxes transformed in the sawmills (AGRESTE, 2022), and the energy wood survey of Environmental statistics ("La consommation de bois-énergie des ménages en 2020," 2024). The temporal continuity of harvest statistics over the study period (1990-2022) is therefore not guaranteed, and has not been studied so far.

To obtain the changes in volume and carbon stocks in forests over the entire country or large regions, a statistical inference method is applied to the ensemble of the forest plots in the region of interest, with plot weights being dictated by the sampling design. The weights used by the NFI are not publicly available to reproduce the calculation, but the accuracy of the mean carbon stock at the national scale has been estimated to be on the order of 4% (REF). At the scale of smaller regions, the high heterogeneity of French forests—where 1-ha parcels can host diverse species, tree sizes, and management histories—leads to inevitable limitation by sampling size. As a result, errors in estimating mean volume, carbon stocks, and their changes over five-year intervals are larger. Error quantification can be accessed using the online tool OCRE (https://ocre-gp.ign.fr/ocre). We present and comment on the results from the French national forest inventory and other sources, including net carbon changes and gross gains and losses at national scale in section 1, and at regional scale in section 2.

This study is based on using the publicly available data originating from the National Forest Inventory and other data sources, processed by CITEPA, the national carbon inventory agency. We analyze how carbon stocks and fluxes causing stock changes in forests have changed between 1990 and 2022, as reflected from these official data. The analysis of the net carbon sink is complemented by studying the evolution of gains from forest growth and losses from natural mortality and harvest. The results are shown at the scale of Metropolitan France and large (NUTS2) administrative regions.

The most recent year for which data are available and were published is 2024. The data are based on the national official report of France to the UNFCCC compiled by the national inventory agency CITEPA (https://unfccc.int/documents/627737). CITEPA provides additional information on regions of Metropolitan France in the National Inventory Document (https://unfccc.int/documents/645100).

The data from different campaigns of the National Forest Inventory which experienced changes in sampling strategies were aggregated by CITEPA. The period 1976 to 2004 has 12-year revisit campaigns performed at department administrative level (NUTS-3) with a rotation across departments every year (plots being fully renewed at each inventory occasion). A non-documented interpolation method was used by CITEPA to elaborate a regional and nationwide overview of forest fluxes. The period 2005 - 2009 was the first version of annual campaigns. The period 2010 to current has a revised version of annual campaigns with 5-year revisits (notion of semi-permanent sampling unit) with multiple updates: identification of harvested trees since 2010, and actual tracking of trees that died since 2015, with re-measurement of circumferences of the living trees, etc. To deal with those methodology changes and data gaps, CITEPA did not use campaigns 1976-2004, but only used NFI data from 2005, provided a point estimate for 1990, and made an interpolation for the beginning of the time series between 1990 and 2005 (see UNFCCC National Inventory Document -2024 report https://unfccc.int/documents/645100 page 261).

The standard forest inventory reports provide mean changes as five-year running averages in order to increase the precision of estimates, yet with an acknowledged risk to lag evolving forest trends or events (Van Deusen, 2011). The estimates for the successive median years indeed vary over time as new samples are re-measured and reveal new changes, with an increasing proportion of annual samples covering the disturbance among five. Thus, only after five samples have been inventoried after a given disturbance, is its impact reported without bias. For instance, if one climate shock in one year is causing large mortality of trees, like for instance, the severe summer drought and massive fires in forests that had not burnt before in 2022, only after 2027 will the 5-year running average reflect the disturbance intensity. However, the use of annual samples in a disturbance context is possible, as 5-year averages are only intended to secure statistical precision.

Second, even this final estimate may still substantially underestimate carbon losses if they come from small-scale (so-called rare) but intense disturbance events occurring in regions where only a limited number of forest plots are sampled. For instance, in 2022, 60 000 ha of forests have been affected by fire events, corresponding to roughly 0.35% of the forest area. With 7000 plots re-measured in 2023, only about 25 plots would capture these events at a national scale. When such events are concentrated in one department, their assessment relies on roughly 80 sampling plots that can be way less than what would be needed to assess carbon losses during the shock year, in particular in the administrative units where large fires or mortality events have occurred. This is a major reason why ongoing research in forest inventory is developing remote sensing based inventory, and disturbance-based design approaches as monitoring options adapted to the current environment (van Deusen 2000, gtr_nc213_014.pdf)

Finally, tree-level data provided at each sample location allows us to quantify natural mortality during the revisit occuring 5 years later, provided that these trees have not been harvested in the meantime (see above). We show in section 2 that natural mortality has increased dramatically over the recent years in different regions, and we will present in section 3 more information about where in France trees have been dying, which species have been dying more frequently, and which size of trees have been more affected over the last years.

**1. Changes in carbon stocks at national scale**

Fig 1 shows the carbon budget of the forest sector in France, with negative values indicating fluxes of CO2 removed from the atmosphere and positive values indicating fluxes of CO2 lost to the atmosphere. The net carbon balance shown using the black curve demonstrates a net gain, implying that forests form a carbon sink all along the period. This net carbon sink increased from 49.3 MtCO2 yr$^{-1}$ in 1990 to reach a peak of 74.1 MtCO2 yr$^{-1}$ in 2008. During this first period, the linear rate of relative increase is 2.45% % yr$^{-1}$. After 2008, the sink rapidly declined and reached a value of 38.8 MtCO2 yr$^{-1}$ in 2022. This large decrease of the sink after 2008 is not linear and follows three stages. From 2008 to 2013, a small decrease of 6.23% was observed, starting with a drop in 2009 consecutive to the passage of the cyclone Klaus. From 2013 to 2017, a rapid and large decrease of 43.05% was observed. After 2017, the sink kept stable to a low value of 37.8 MtCO2 yr$^{-1}$. Since 2013 is the median year of the first revisit campaign with the new semi-permanent sampling protocol established in 2010, it cannot be ruled out that the sink decrease seen in Fig. 1a could be a transient methodological artifact.

Looking now at the different components of the carbon sink, we examine separately the gains and the losses. The annual gains are dominated by the growth of established forests (light green). The growth of those established forests has evolved roughly in parallel with the net carbon gain (R2 =0.74, p<0.05) from 1990 to 2008, although the increase and decrease of growth before and after the maximum of the net carbon sink in 2008 only explains 47.70% of the variation of the net sink. Interestingly, the growth of forests reached a maximum in 2014, then decreased slightly by 5.51% to reach a minimum in 2017 and remained stable or slightly increased between 2017 and 2022 (but see Hertzog et al. 2025 for a recent detailed scrutiny with inventory data) (Hertzog et al., 2025). The creation of new forests from areas converted from another land use type to a forest (dark green), accounted for 9.3 ± 3.4% of the total annual gain, that is a CO2 removal from the atmosphere of 13.3 Mton CO2 yr$^{-1}$. The carbon sink of these new forests steadily declined from 19.49 in 1990 to 10.05 in 2022, reflecting a slowdown in the area of new forests each year during the whole period.

The third component of the net carbon gain is the accumulation of soil and litter, a term which is not measured but modeled and has a very large uncertainty. For the established forests (light green bars in Fig 1a) in the inventory makes a neutrality assumption that there is no storage in soil carbon. The only increase in the soil taken into account is due to the constitution of stocks in new forests. In contrast, measurement of soil carbon change at 120 long term forest monitoring sites in France (mature forests with little management) suggests a large rate of increase of soil carbon of 1,28 ton CO2 per ha (Jonard et al., 2017). On the other hand, meta-analysis shows that after clearcut and intensive management, up to 20% of the top soil carbon is lost to the atmosphere, partly neutralizing the role of a carbon sink in forest soil. This loss of carbon from management activities is not included in the National Inventory of forest carbon.

The fourth component of carbon gain is the storage of carbon in wood products in long-lived pools such as construction materials, furniture and landfills. This term is not measured but calculated using harvested timber input data and sectoral models of wood transformation and lifetime in different pools. Wood products are a carbon sink, which mean that their mass increased over time, but this sink represents a very small component (4.9±2.60%) of the total carbon gain, and it decreased from 5.40 Mton $CO_2$ $yr^{-1}$ in 1990 to 1.09 Mton $CO_2$ $yr^{-1}$ in 2022. This implies that despite increasing carbon storage in wood products being increasingly recognized as a climate mitigation option, the efficiency of this sink has decreased in France, probably reflecting an increasing share of wood harvested being transformed to fuel wood for residential heating and power production, and a lower quality of wood products to secure low prices which causes more wood thrashing in landfills. It also appears low as compared to the in situ forest C sink at the European level (EC 2021, https://environment.ec.europa.eu/strategy/forest-strategy_en).

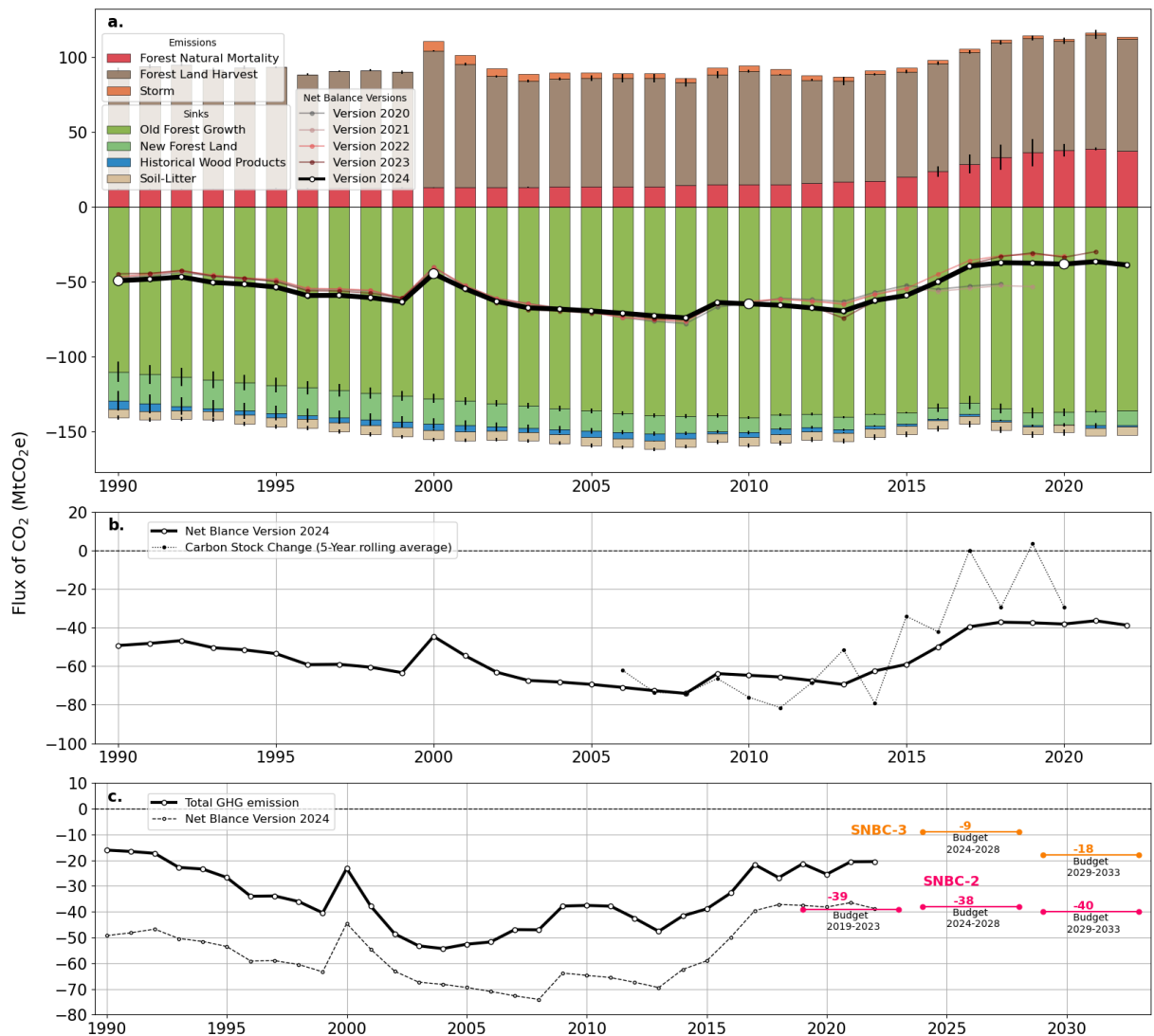

**Fig 1. (a)** Forest carbon gains and losses and net change in Metropolitan France. Carbon sequestration or CO2 removal from the atmosphere annually from the growth of living trees in established forests, the creation of new forest, the increase in soil carbon and litter and storage in wood products are based on measurement campaigns from the national inventory updated each year and calculations using the life cycle of products for wood product pools. Carbon gains are reported as negative numbers in the figure. Carbon losses from harvest, mortality and large disturbances from storms are reported as positive numbers. The net change is the thick black curve from the latest national report of France, the grey curves are previous estimates published in previous years. **(b)** Comparison between the net carbon sink reported by France to UNFCCC and a direct year on year change of stocks from ≈ 7000 forest plots measured each year during each NFI campaign (dotted). The variability in this curve reflects noise in the statistical sampling to have a national stock change estimate **(c)** Net carbon sink of forests compared to the national trajectory (Stratégie Nationale Bas Carbone, SNBC) for the whole land use sector from SNBC-2 revised for the current period and the two future five-years periods until 2030 and the SNBC-3 for the same two future periods with the land use sector objective targets being revised down.

The annual losses shown as positive numbers in Fig. 1a are the sum of harvest removals (brown bars), natural mortality encompassing causes such as fires, diseases, insects, drought, frost, small storms, and the continual death of individual trees from competition between individuals in dense forests (red bars), and large disturbance events mainly from two extra-tropical cyclones reaching explosive development rates with winds gusts on the order of 200 km hr-1 : Lothar and Martin in December 26-29 1999 and Klaus in January 23-24, 2009 (orange bars) (U. Ulbrich, A. H. Fink, M. Klawa, J. G. Pinto, 2001). The main factor of carbon loss remains harvest, 75.8±5.1 Mton $CO_2$ $yr^{-1}$ . The removal of wood by harvest offsetted about half of the annual growth during the whole period from 1990 to 2022. More precisely, during the first period of 1990-2008 when the net carbon sink increased, harvest offsetted 58.5± 7.8 % of the growth. During the second period from 2008 to 2017 when the carbon sink quickly decreased, harvest offsetted 52.0± 2.7% of the growth. During the more recent period from 2017 to 2022 with a stable net carbon sink at a low value, harvest offsetted 55.4 ± 1.4 of the growth. The stability of this ratio indicates that there has been no long term increase of harvest pressure in the French forests over the last 32 years. Yet there was a strong impact of a transient increase of harvest during the period when the sink decreased in 2013-2017, which is discussed below together with mortality changes.

Moreover, during the entire period, variations in harvest only explain 26.0 % ($R^2$ = 0.26) of the changes in the net carbon sink, thus much less than the variations of growth. We also see in Fig 1a the effect of salvage harvest consecutive to massive tree death caused by the two explosive cyclones events. *Lothar* and *Martin* in December 1999 laid down 300 M trees, (John Abraham, Fouad Bendimerad, Agnete Berger, Auguste Boissonnade, n.d.) 3% of the national total, mainly in the Atlantic region where *Martin* made landfall, and in Normandy and the northern part of France crossed by Lothar. This loss of 176 million of $m^3$ was equivalent to three years of normal harvest (Gardiner et al., 2013). A fraction of the uprooted and broken dead trees were salvaged and sold later by the wood industry, showing up as a peak of harvest in 2000 and 2001 in Fig. 1a. A smaller increase of harvest is also observed after the cyclone *Klaus* which made landfall in 2009 near Bordeaux and laid down 42 $Mm^3$ (95% confidence interval of ± 40) of trees, mainly in the *Les Landes* plantation forest (Pawlik et al., 2022). Mortality from extreme winds during these cyclones affected large contiguous areas of forests in specific regions, causing massive tree losses, while small windblown events during other years may be undersampled by the inventory. . Analysis of airborne photos, ground surveys and models made it possible to estimate separately the immediate carbon losses from biomass destroyed during the cyclones and the legacy carbon losses from non harvested dead trees and branches which decomposed slowly on the ground several years after the passage of each cyclone (IFN, 2009) ("Les tempêtes de décembre 1999 - Bilan national et enseignements," 2003). The net $CO_2$ emissions caused by the cyclones (orange bars in Fig 1a) represent 5.65 % of the total carbon loss during the year of the event, with a legacy effect of 5.02% of the total carbon loss during the following five years.

Besides the two peaks of carbon losses when the cyclones struck, the second largest cause of carbon loss after harvest is natural mortality (red bars). Between 1990 and 2001, natural mortality was stable or increased very slowly, and represented 13.1 ± 0.7 % of the total loss, equivalent to 15.3 ± 0.9% of the loss due to harvest. After 2013, mortality suddenly accelerated and has been multiplied by three between 2013 and 2017. During this critical period which saw a large and fast decrease of the net carbon sink, the increase of mortality explained 51.5 % of the sink decrease and a coincident increase of harvest explained 48.5%.

Note that this increased harvest signal could reflect salvage harvest of recently dead trees after mortality events. Intriguingly, the crisis period of dying trees in the French forests appears to have started before the severe droughts and heatwaves recorded in 2018, 2022, 2023, even though 2015 was marked by a strong water deficit and hot temperatures in summer (Orth et al., 2016). This sudden change should be considered with caution as it is coincident with a change in protocol in 2015. Between 2018 and 2022, despite more frequent and more severe summer heatwaves and droughts, natural mortality has remained large but it has not increased further (Fig. 1a). Over this recent period, natural mortality represented 48.7% of the harvest, an unprecedented loss of wood for the economy. Mortality could be an even greater fraction of harvest since salvage harvest or sanitary cuts over insect-affected areas follows mortality events and will be classified as harvest by the inventory. Unfortunately, the data collected by the National Forest Inventory do not allow us to separate salvage from normal harvesting to date, as it would require a faster remeasurement of first-visit forest inventory plots, the issue being currently under scrutiny.

Importantly, deadwood has likely increased over time in response to recent mortality events, but changes in deadwood carbon are not measured or estimated by the inventory in this edition, except after the two cyclones. Possibly, deadwood carbon is now increasing on forest floors, which results in a transient carbon accumulation in the French Forests but will give a legacy $CO_2$ emission in the near future when recent dead woody debris will decay.

Since each year, France sends a national communication to the UNFCCC to report its official emissions and sinks of greenhouse gases to the United Nation Convention on climate change, based on current inventory data from two years ago, we collected data from successive reports published from 2020 to 2024 from the UNFCCC website. Significant changes were identified between the successive reports, with the 2021 edition underestimating the recent $CO_2$ sink decrease, and the 2023 edition overestimating this decrease compared to the latest 2024 edition (black thick line). These changes reflect the inclusion of new sites measured by the inventory each year, thus impacting thus the moving window estimates. They are also sensitive to regular changes in the protocols or conventions for computing forest carbon accounting estimates (Van Deusen, 2011) (Francis A. Roesch, James R. Steinman, and Michael T. Thompson, n.d.). Note that climate shocks leading to abrupt losses of $CO_2$ in one year will be smoothed in time by the 5-years reporting window of the inventory. Extra $CO_2$ losses from wildfires were included using a specific yearly approach.

Fig 1b compares the net forest carbon sink reported by France to UNFCCC with a direct year on year change of carbon stocks from ≈ 7000 forest plots measured each year during each NFI campaign calculated by the NFI research laboratory (LIF Nancy) (dotted). The annual variability in the stock change curve reflects noise in the statistical sampling to produce a national stock change estimate. Nevertheless, we can clearly see that the direct stock change method gives a carbon sink about two times smaller than the UNFCCC report in recent years, likely reflecting the fact that data from the latest inventory campaigns were not yet included in the CITEPA estimate. This result clearly shows that the current report to UNFCCC very likely overestimates the real carbon sink in the French forests.

France has adopted a national law on carbon neutrality (Stratégie Nationale Bas Carbone - SNBC) which defines and revises carbon emissions budgets for different sectors for successive five year periods. For the land use, land use change and forestry sector (LULUCF), the first SNBC-1 published in 2015 did not have any specific target. While other emitting sectors have specific targets, the SNBC-2 had a target sink for the LULUCF of -39 Mton CO2e yr-1, revised to -43 Mton CO2e yr-1 in 2024 for the first budget 2019-2023, which is about the magnitude of the forest sink alone (Fig. 1c) (French Government, 2020). However, this target defined in CO2 equivalents also includes emissions and absorptions in the LULUCF sector outside forest, mainly in croplands and grasslands, which have emitted around 9 Mton CO2e yr-1 in the recent years. Hence the current objective of SNBC-2 for the LULUCF sector as a whole has not been met, and further goals in 2024c-2028 and 2029-2033 shown in red in Fig 1c are unlikely to be achieved. The government reduced the ambition in the LULUCF sector and the SNBC-3 has proposed a more modest sink goal of -9 Mton CO2e yr-1 for 2024-2028 and - 18 Mton CO2e yr-1 for 2029-2033 (French Government, 2024). Will the current forest sink reported to UNFCCC remains stable, the objective would be met. If we use the direct stock change approach to estimate the sink, then the SNBC3 objective will not be achieved and France is already a net source of CO2e in the LULUCF sector (Fig 1b), then If it further reduces in view of increased disturbances, reducing growth (Hertzog et al. 2025, and the long-delayed effect of forest renewal and afforestation policies (https://agriculture.gouv.fr/francerelance-le-renouvellement-des-forets-francaises), SNBC-3 goals set for the next decade may prove challenging to reach.

**2. Changes in carbon stocks at regional scale**

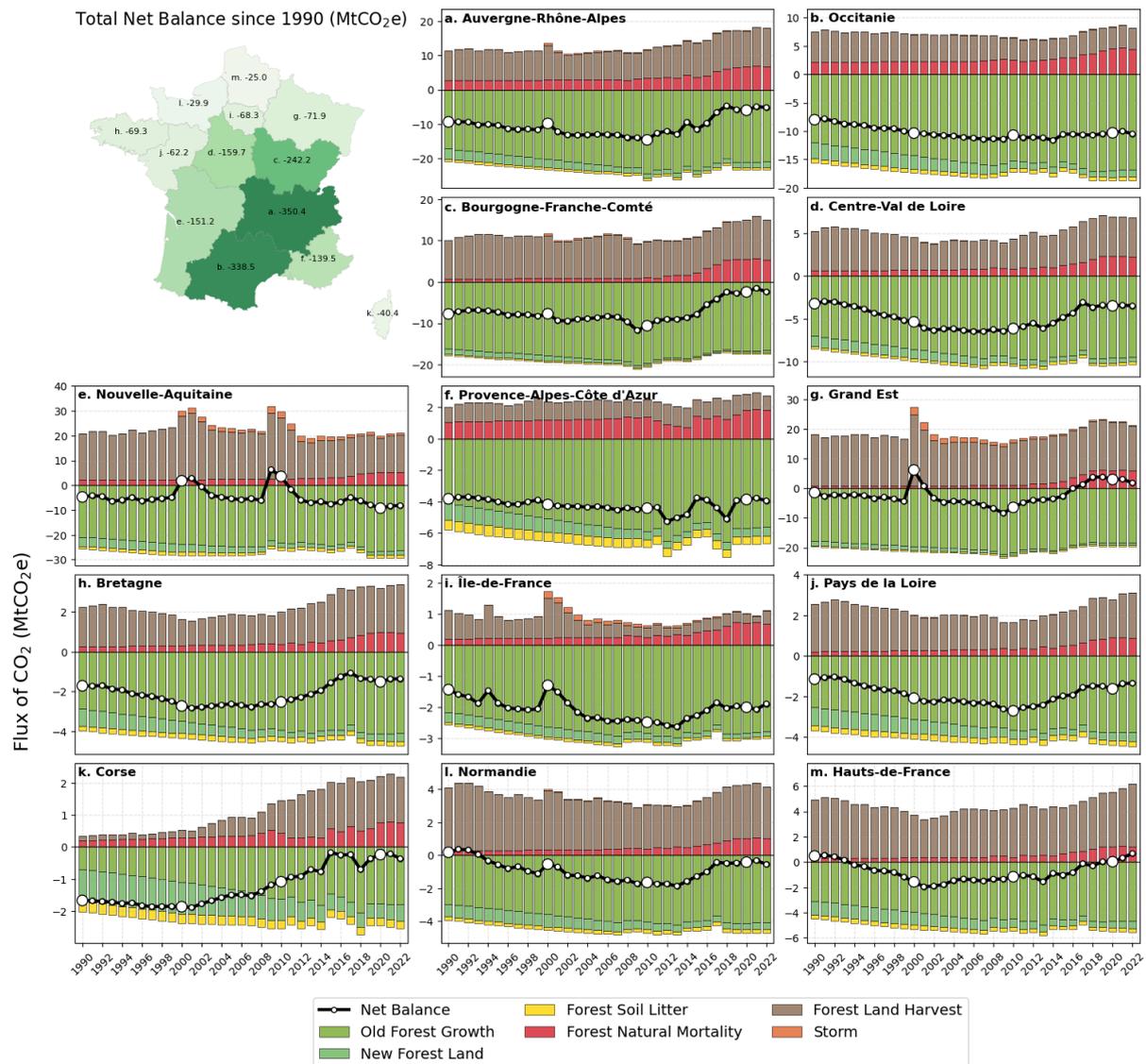

**Fig. 2.** Same as Fig. 1a but for 13 administrative regions of France indicated in the map.

In this section, to gain insight on the contribution of each region to the nation-wide reduction of the carbon sink and increase of carbon losses, we analyze regional trends. Fig 2 shows carbon gains and losses of the 13 different administrative regions (NUTS2) in France. Carbon sequestration or CO2 removal from the atmosphere annually from the growth of living trees, the creation of new forest, and increase in soil carbon and litter are based on annual campaigns from the national inventory. They are reported as negative numbers in the figure. Carbon losses from harvest, mortality and large disturbances from storms causing windthrown events with massive tree killing are reported as positive numbers. Note : changes in wood products pools are not reported per region but they are shown at national scale in Fig 1.

We observe contrasted trends across the 13 regions of Metropolitan France in Fig. 2. The *Hauts de France* and *Grand Est* forests have become net emitters of CO2 to the atmosphere in recent years, meaning that they no longer contribute to mitigate climate warming. Other regions saw a strong decline of their CO2 removal rates, which approached zero like in Normandy, Corsica and Bourgogne-Franche Comté. There is no straight relationship between management intensity and the decline of the CO2 removal by forests in the west and the north of France. For instance, the intensively managed plantation of maritime pine in *Les Landes* (Fig. 2e), yet largely reafforested after the storms of 1999 and as such constituting a carbon sink, shows an increasing CO2 sink while intensively managed broadleaved and conifer forests in Northern and eastern France show a strong decline (Fig. 2g and 2m). The southern regions of Occitanie (Fig 2b) and Provence (Fig 2f) and where forests are not intensively managed, show a sustained CO2 sink. In contrast, Corsica went from a large CO2 sink to nearly zero because harvest increased enormously in that region (Fig 2k) (Suvanto et al., 2025).

The net carbon sink reached a peak earlier than the national average (2008) over Britany (Fig 2h), Corsica (Fig 2k), Centre Val de Loire (Fig 2d), and Hauts de France. The steepest declines of the sink during the period when the national sink declined from 2008 to 2017 have been observed at a rate of 8.2 ± 2.0 Mton CO2 yr$^{-1}$ ( 8.7 ± 9.2% per year). The signature of the two cyclones is more apparent at the regional level than at the national scale (Fig. 1). *Lothar* and *Martin* in 1999 caused the strongest decreases of the carbon sink in Nouvelle Aquitaine, Grand Est and Ile de France followed by a rapid recovery in the four consecutive years. Klaus in 2009 affected mainly Nouvelle Aquitaine. In each of these impacted regions, after the passage of the cyclones, a transient increase of harvest reflecting salvage wood recovery is observed. In Provence and Corsica which represent usually the most frequently burned regions (Vallet et al., 2023), extreme fire years of 2009 and 2016-2017 respectively cause a loss of carbon sinks of about 28.40%, still followed by recovery within 5 years. Note that extreme fires in 2022 in Nouvelle Aquitaine (23163 ha of forest area burned) are not yet seen in the latest data available from the national inventory and will likely be included as a large carbon loss in that region, estimated by an independent study at 6.23 Mton CO2 yr$^{-1}$ (Vallet et al., 2023).

Harvest increased dramatically in Corsica after 1998, and Brittany and Hauts de France, Pays de Loire after 2013, in parallel with the rise of mortality. It is impossible to assess whether this increase of harvest is a consequence of sanitary cuts of dead trees or if it is due to other factors. Interestingly, the ratio of harvest-to-growth is large in northern regions (average 0.86 ± 0.32) where most forests are accessible production forests being intensively managed and southern and mountain regions (average 0.41 ± 0.20) where lack of access roads and terrain limit the extraction of wood. The lowest harvest-to-growth ratio was found in Île-de-France (0.21), Provence-Alpes-Côte d'Azur (0.22), Occitanie (0.31) and the highest one for the intensive plantations of Les Landes, covered mainly by maritime pines with a typical harvest rotation of 20 years.

We can also see in Fig 2 that natural mortality increased in all the regions almost coincidently around the year 2013, but the magnitude of the increase differs strongly between regions, with the smallest mortality increase in Provence-Alpes-Côte d'Azur (6.46% per year after 2013) and Occitanie (7.15% per year) and the highest in Corsica (10.40% per year) and Auvergne-Rhône-Alpes (9.28% per year). In Grand Est affected by droughts and recently by massive bark beetle attacks on spruce forests after 2018, mortality increased by *% per year after 2015. In all the regions, mortality showed a sharp increase between circa 2013 and 2017-2020 and remained stable thereafter, despite severe drought conditions in 2022 and 2023, which may be revealed as increased carbon losses in subsequent inventory campaigns yet to be performed and published. The reasons why mortality increased in 2013 across a majority of regions is still unclear. Previous heat waves like the one of summer 2003 did not seem to cause a large increase of mortality, and the warming rates have been high over the last decades but did not suddenly accelerate in 2013. One possible explanation is that the harvests are being estimated by the the national forest inventory since 2010 using semi-permanent plots, while they were estimated from external and indirect data sources before (Enquête Annuelle de Branche, et enquête bois énergie, see above). Also, semi-permanent NFI plots may detect more mortality events since the protocol change, and its extension to mortality in 2015. Changes in the NFI protocols have been implemented to obtain strong gains in the accuracy of forest fluxes, in a monitoring perspective requested by public policies, in view of ongoing climate change(Hervé et al., 2014). A major consequence however lies in the difficulty to obtain a robust temporal view of forest changes concomitant to accelerated climate warming in France, urging an unprecedented effort for homogenizing past data and establishing a firm retrospective reconstitution of forest carbon in France.

## 3. Changes in tree mortality across regions, species, and height classes

More insight into tree mortality across different regions, species, and tree size classes was gained by analyzing the NFI data collected during first visits between 2010 and 2018, with re-visits occurring five years later from 2015 to 2023. Following a revision of the sampling protocol in 2010, each tree in the sample (619,496 trees from 50,012 forest plots) was individually monitored to determine survival or mortality over the five-year re-visit period. This approach improved the accuracy of tree mortality estimations.

We quantified tree mortality using two metrics: volume-based mortality, measured as the cubic meters of wood lost per year and hectare forested area (m³ yr⁻¹ ha⁻¹), and stem-based mortality, expressed as the percentage of censusable stems lost per year (%-stems yr⁻¹). Mortality ($M$) was calculated using the formula (Kohyama et al., 2018): $M = [1 - [N_{t1} / N_{t0}]^{[1/t]}] \times 100$, where $N_{t1}$ represents the number of trees alive at the first visit, $N_{t0}$ is the number of individuals that survived between visits, and $t$ is the time interval between visits (five years). To focus on natural mortality, harvested trees were classified as survivors, ensuring that the estimated rates reflected mortality independent of logging. However, it is important to note that large disturbances, such as storms or bark beetle outbreaks, are often followed by rapid salvage logging. Due to the five-year revisit interval, our methodology does not capture these short-term responses, meaning the reported mortality rates likely underestimate actual natural mortality.

To assess uncertainty, we applied a bootstrapping approach. For each year and group (e.g., region, height class, species), we resampled the dataset and calculated the mortality rate across multiple iterations. The final estimate represents the mean mortality rate across all bootstraps, providing a robust measure of variation.

Figure 3 presents the spatial distribution of volume-based mortality rates due to natural causes, mapped using a hexagonal grid. The data reveal a clear increase in tree mortality across France, particularly after 2018, with the most pronounced effects in the northeastern regions. When aggregated over larger eco-regions ("Grande Région ECOlogique") (IGN, 2024), Figure 4a shows that all regions shared a similar mortality rate from 2015 to 2017 of approximately 0.25 to 0.5 m³ yr⁻¹ ha⁻¹. However, in 2018, mortality rates started to surge predominantly in the Jura (a 10-fold increase from 2015 to 2023), Vosges (6-fold), and Grand Est (4-fold), in likely response to the extremely hot drought occurring in central Europe (Schuldt et al., 2020). In mountainous regions such as the Alps, Massif Central, and Pyrénées, mortality still rose less drastically but continuously over time, reaching rates of approximately 0.75 $m^3\ yr^{-1}\ ha^{-1}$. The mortality rates along the Mediterranean and Atlantic coasts are generally smaller than in other regions (below approximately 0.5 $m^3\ yr^{-1}\ ha^{-1}$) but still showed 2- to 3-fold larger mortality rates in 2023 compared to 2015.

Figure 4b highlights species-specific mortality trends, revealing that Norway spruce, European ash, and silver fir accounted for the highest volume-based mortality, with a sharp increase after 2018, reaching rates up to 40 $m^3\ yr^{-1}$. The rise in mortality among Norway spruce and silver fir is closely linked to both direct and indirect effects of climate change. A 2019 study on the Vosges forests attributed increasing mortality rates in both species to reduced water availability and severe drought events (Piedallu et al., 2022). Norway spruce, in particular, is highly susceptible to drought stress (Arend et al., 2021), which weakens trees and makes them more vulnerable to bark beetle infestations (Hlásny et al., 2021). These outbreaks have intensified in recent years, as climate change has created more favorable conditions for beetle populations—warmer winters improve their survival rates, while hotter summers accelerate their life cycle, allowing additional generations to develop (Hlásny et al., 2021). The dramatic rise in European ash mortality (a 16-fold increase from 2015 to 2023) is primarily driven by the spread of the fungal pathogen *Hymenoscyphus fraxineus*, which can cause mortality in up to 85% of infected ash trees (Carroll and Boa, 2024).

Unlike defoliating insects, which typically weaken but rarely kill trees, bark beetles bore through the bark of their hosts, disrupting the flow of sap and ultimately causing tree death. While beetles have long been endemic to Europe, they have not been a major source of tree mortality until recently, when successive summer droughts weakened trees' natural defenses. The resulting outbreaks have had widespread consequences, affecting not only France but also Belgium, Germany, Austria, and the Czech Republic (Hlásny et al., 2021). In the Czech Republic, severe beetle infestations have transformed forests dominated by spruce from a net carbon sink—absorbing 10 million tons of $CO_2$ per year—into a net carbon source, emitting an equivalent amount annually since 2018 (Arend et al., 2021; UNCC, 2020).

Moreover, mortality rates have risen significantly across all major tree species in France, including European beech, oaks (pedunculate, sessile, and downy oak), common hornbeam, and sweet chestnut (e.g., chestnut is also known to suffer from pathogen attacks (Jung et al., 2018) and beech suffers from extreme droughts (Leuschner, 2020)). From 2015 to 2023, mortality in these species has at least doubled, and in some cases, tripled. When looking at %stems-based mortality, ash, chestnut, and spruce have substantially larger rates than other species, highlighting the actual decline of these populations. This trend underscores the growing pressure that Europe's forests face under climate change (Senf et al., 2020), emphasizing the urgent need for adaptive management strategies and targeted policy measures to safeguard the critical ecosystem functions these forests provide.

It is important to note that volume-based mortality rates are particularly relevant for understanding carbon sink dynamics. However, these rates are inherently biased toward the mortality of large trees and provide little insight into the survival of younger tree generations. Figures 4c and 4d illustrate trends in both volume-based and stem-based (%-stems) mortality rates. While small trees (under 10m in height) exhibit lower volume-based mortality rates compared to taller trees (over 25m), they have significantly higher stem-based mortality rates. This discrepancy reflects natural forest dynamics, where smaller trees primarily die due to competition for light rather than external disturbances (Westoby, 1984). Nonetheless, the percent-stem-based mortality rate indicates that mortality has increased across all height classes, highlighting the broad-scale impact of recent environmental stressors. Given the current species mix, this could limit the number of mature trees available as well as the carbon sink in the coming decades. However, shifts in species composition, with the establishment and spread of more resilient or better-adapted species, may influence long-term forest dynamics and carbon sequestration potential (Wessely et al., 2024).

Although large trees have lower mortality rates compared to smaller trees, volume-based mortality highlights their outsized contribution to carbon stock changes. This aligns with the impacts of storms discussed earlier, as larger trees are more vulnerable to windthrow (Seidl et al., 2017), making storms significant drivers of carbon loss. Additionally, large trees are highly susceptible to drought stress due to their expansive canopies, which create a high evaporative demand and generate strong pressure gradients from soil to atmosphere (Fensham et al., 2019) (Bennett et al., 2015) (Fensham et al., 2019). Under extremely dry conditions, negative pressure within the xylem (the water transport vessels of a tree) can lead to the formation of gas bubbles, disrupting water flow and causing leaves and branches to desiccate and die. This loss of foliage reduces carbon assimilation, weakening the tree's ability to sustain metabolic functions. As structural integrity declines, the tree becomes even more prone to windthrow and secondary disturbances such as bark beetle infestations. Since tree height is, however, also a factor of vulnerability to water stress (Koch et al., 2004), mortality rates should increase with tree height under the climatic drought pressure regularly experienced over the past years. In view of the relationship found in Fig 4d, density-dependent mortality exacerbated in young forests is therefore interpreted to dominate this response. Further research is here requested to elucidate this response, that may stem from an insufficient cover of larger height classes (beyond 30 to 40 meters). The sharp rise in volume-based mortality over the past decade is a likely consequence of repeated extreme droughts in 2018, 2020, and 2022 rather than storm events.

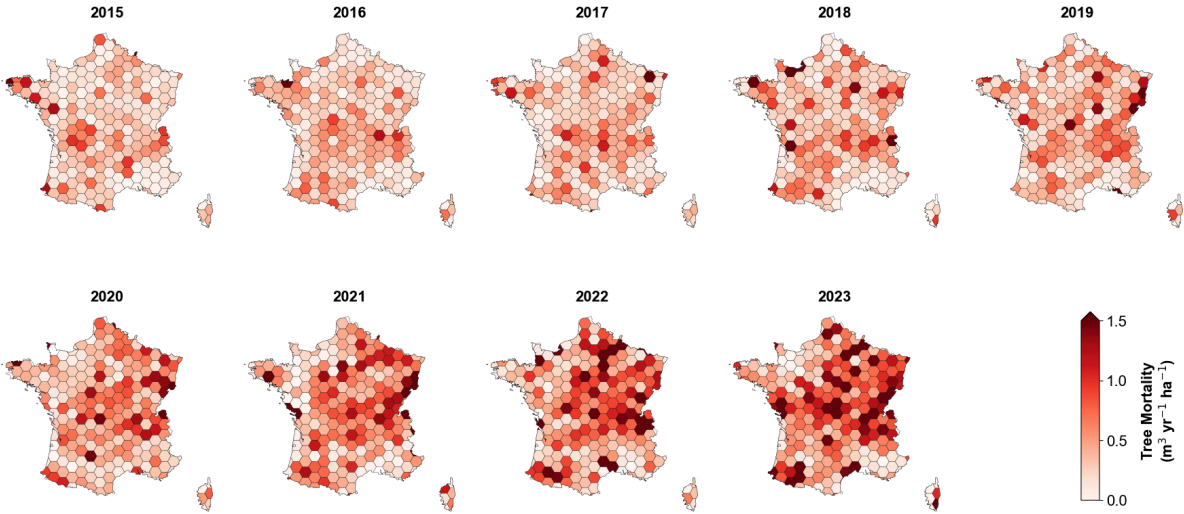

Fig 3. Spatial distribution of volume-based tree mortality rates across France from 2015 to 2023. Mortality rates increased substantially after 2018, particularly in northeastern regions, reflecting climate-induced stressors and pathogen outbreaks.

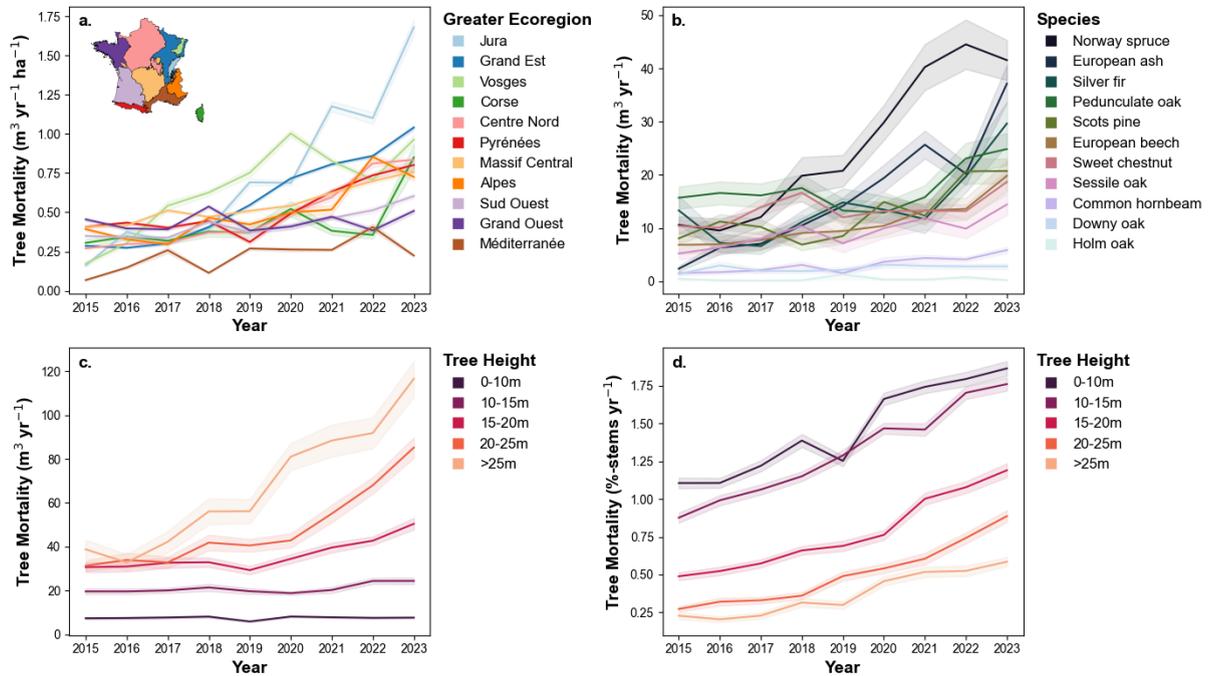

Fig 4. Trends in tree mortality across different categories. (a) Volume-based mortality trends by eco-region, showing a sharp rise in the Jura, Vosges, and Grand Est regions. (b) Volume-based mortality trends across species, with Norway spruce, European ash, and silver fir experiencing the highest volume losses. (c-d) Mortality trends by tree height class, illustrating the differential impacts of forest disturbances on large versus small trees (taller trees contributing more to carbon losses despite lower %-based mortality rate).

**4. New tools to monitor forests using satellite and ground observations**

Ground-based observations, such as the French National Forest Inventory surveys, provide high-quality data for monitoring forest dynamics at the national or regional scale (Fig 1, 2). However, these data have several limitations. First, the temporal resolution for forest flux measurements is low, as the revisit interval between successive measurements spans five years, a trade-off between detecting substantial changes and increasing the resolution with respect to previous practices (around 12 years in the former approach). Also, samples are renewed every year, allowing to quickly detect changes in at least state variables. This low-frequency sampling of fluxes nevertheless hinders the timely detection of sudden events affecting e. g. the growth or mortality of French forests, such as the extensive wildfires of 2022, or tree mortality caused by storms and pest outbreaks. A perspective here may be to panelize the revisits across the 5 successive years, yet at the sake of design simplicity. Second, although NFIs are designed to be statistically representative, they do not monitor forests at a fine resolution. This introduces uncertainty if the statistical sampling fails to account for spatial heterogeneity, and bias if forest events are disseminated across space to the point where they can be qualified as "rare" events, locally. Additionally, traditional NFIs have no capacity for high-resolution, small-scale monitoring, for the reason that they have been designed to support national forest policies. However, higher-resolution is increasingly crucial in France, where forest parcels are often small and managed by diverse owners with distinct practices. This aspect is at the origin of model-based mapping and monitoring based on remote-sensing products, also is a cause for enhanced design-based forest inventory development (so-called multi-source inventory) where remote sensing products only play an auxiliary role for forest mapping and increase in estimation precision.

In recent years, remote sensing (RS) technology has undergone significant advancements, particularly with the development of new satellite sensors that provide vast amounts of data for forest monitoring. Among them, the Sentinel-2 (S2) mission, part of the European Space Agency's (ESA) Copernicus Earth Observation Program, captures multispectral imagery at a 10-meter spatial resolution with a revisit interval of approximately 6 days in France, making it a valuable tool for detecting changes in vegetation and canopy structure. Similarly, Sentinel-1 (S1), another component of the Copernicus program, provides synthetic aperture radar (SAR) imagery at 10-meter resolution, operating independently of weather conditions and daylight, thus enhancing forest monitoring capabilities.

Beyond optical and radar remote sensing, LiDAR technology has emerged as a powerful tool for forest structure analysis. LiDAR sensors measure the three-dimensional structure of forests by emitting infrared laser pulses and recording their reflection from different canopy layers. These measurements provide critical insights into forest height, biomass, and structural complexity. Since 2018, NASA's Global Ecosystem Dynamics Investigation (GEDI) mission (Dubayah et al., 2020) has delivered sparse but highly accurate data on forest vertical structures across the globe, enabling unprecedented assessments of forest height and canopy density. Specifically, canopy height, one of the most straightforward metrics derived from LiDAR measurements, has shown strong correlations with key ecological indicators such as biomass, biodiversity, and forest health (Dubayah et al., 2020; Torresani et al., 2023). Furthermore, airborne LiDAR missions, such as those conducted by national forest agencies, complement spaceborne LiDAR data by providing denser coverage (Coops et al., 2021) at finer scales like the French LiDAR HD program initiated by the IGN.

Over the past decade, advances in artificial intelligence (AI) and machine learning, particularly deep learning models (LeCun et al., 2015), have transformed satellite remote sensing research. These frameworks can efficiently process large-scale datasets and are particularly effective with unstructured data such as images or sound. This makes them perfectly adapted for the fusion of multiple remote sensing datasets, such as GEDI LiDAR data with Sentinel-1 and Sentinel-2 imagery, to generate continuous, high-resolution maps of forest structure and biomass. The past 5 years have seen an increasing number of studies using these tools to derive such maps globally (Lang et al., 2023) ; (Pauls et al., 2024) ; (Tolan et al., 2024), at continental scale (Liu et al., 2023), or national (Fayad et al., 2024; Liu et al., 2023; Schwartz et al., 2023) and regional scale (Favrichon et al., 2025; Schwartz et al., 2024).

In France, Schwartz et al. (2023) used GEDI, Sentinel-1, and Sentinel-2 data with a U-Net model (Ronneberger et al., 2015) (Ronneberger et al., 2015), a deep learning approach, to generate a 10 m resolution forest height map for 2020, covering the entire metropolitan territory (Fig. 5a). This map demonstrated high accuracy compared to NFI height measurements, with a mean absolute error of 2.94 m, and outperformed all previously available datasets for France. It enables a detailed understanding of French forest structure at the stand level, as visible in Fig. 5b. Furthermore, leveraging allometric equations from NFI data, the authors produced wood volume and biomass maps at 30 m resolution, offering a valuable snapshot of the carbon stored in French forests in 2020.

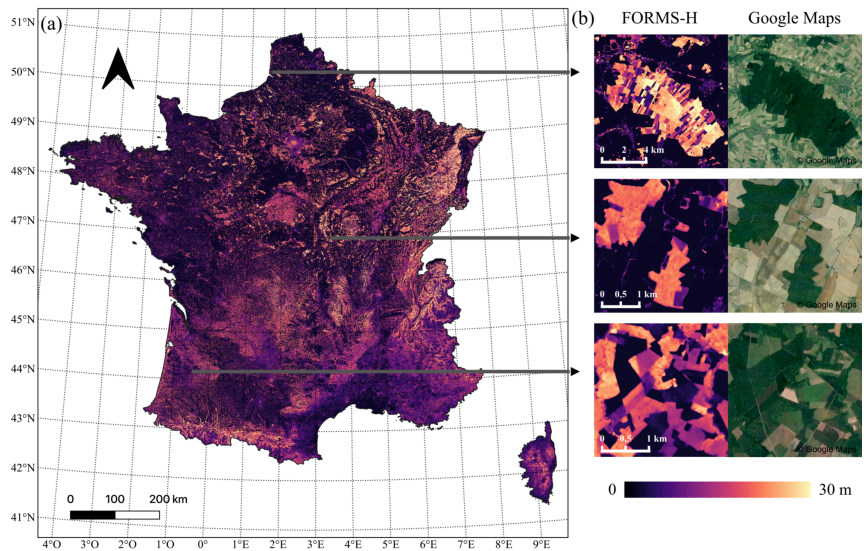

Fig 5: (a) Tree height map of France at 10 m resolution for the year 2020. (b) Examples at three different locations of height prediction (left) with the corresponding Google map images from 2020, 2018, and 2019 (right). Brighter colors indicate higher heights. Figure from Schwartz et al. (2023)

Satellite and AI-based methods for generating forest height and biomass maps have shown relatively high accuracy, particularly in temperate regions like France. However, significant uncertainties remain, especially in biomass estimation. Biomass maps rely on allometric equations fitted by broad forest categories or biomes like in Schwartz et al. (2023) where two equations were used depending on the leaf type of the trees (broadleaf/needleleaf). However, within these broad categories, biomass is influenced by various factors beyond height, including climate conditions, tree species, forest management practices, and tree cover. In addition, volume could be more accurately reconstructed than biomass from remote sensing, given large variations of wood density across species and forest types. These complexities make satellite-based biomass maps less reliable than the height maps they derive from. Future models should incorporate these additional variables into biomass predictions or try to map biomass from direct measurements rather than using height as a proxy.

Forest carbon budget monitoring, meaning an accurate tracking of biomass losses and gains, is crucial to follow climate-related policies and reduce greenhouse gases emissions. With their frequent updates, satellite-based solutions enable this monitoring, and several projects are already operational, including forest cover loss detection (Hansen et al., 2013), tropical forest degradation and deforestation tracking (RADD alerts from(Hansen et al., 2013; Reiche et al., 2021); TMF dataset from (Vancutsem et al., 2021), and clearcut monitoring in France through the SUFOSAT project (Mermoz et al., 2024). The increasing precision of new data and models will soon allow detecting growth signals as well as forest losses to estimate forest carbon uptake. However, few models have successfully generated consistent height time series validated with external data. In France, building on the study presented in Fig. 5, the same authors attempted to address this challenge by developing a framework for predicting height annually in a consistent manner. Fig. 6 provides an initial look at these time-series, where forest growth is visible year to year. In maritime pine plantations (Fig. 6a) of the Landes forest (southwestern France), forest parcels clearly show growth between 2018 and 2024, while clearcuts, represented in red, are also distinct. In contrast, for mature forests with more complex growth dynamics, changes are less visible, and biomass accumulation in tree woody mass may be overlooked, particularly for deciduous oaks (Fig. 6d).

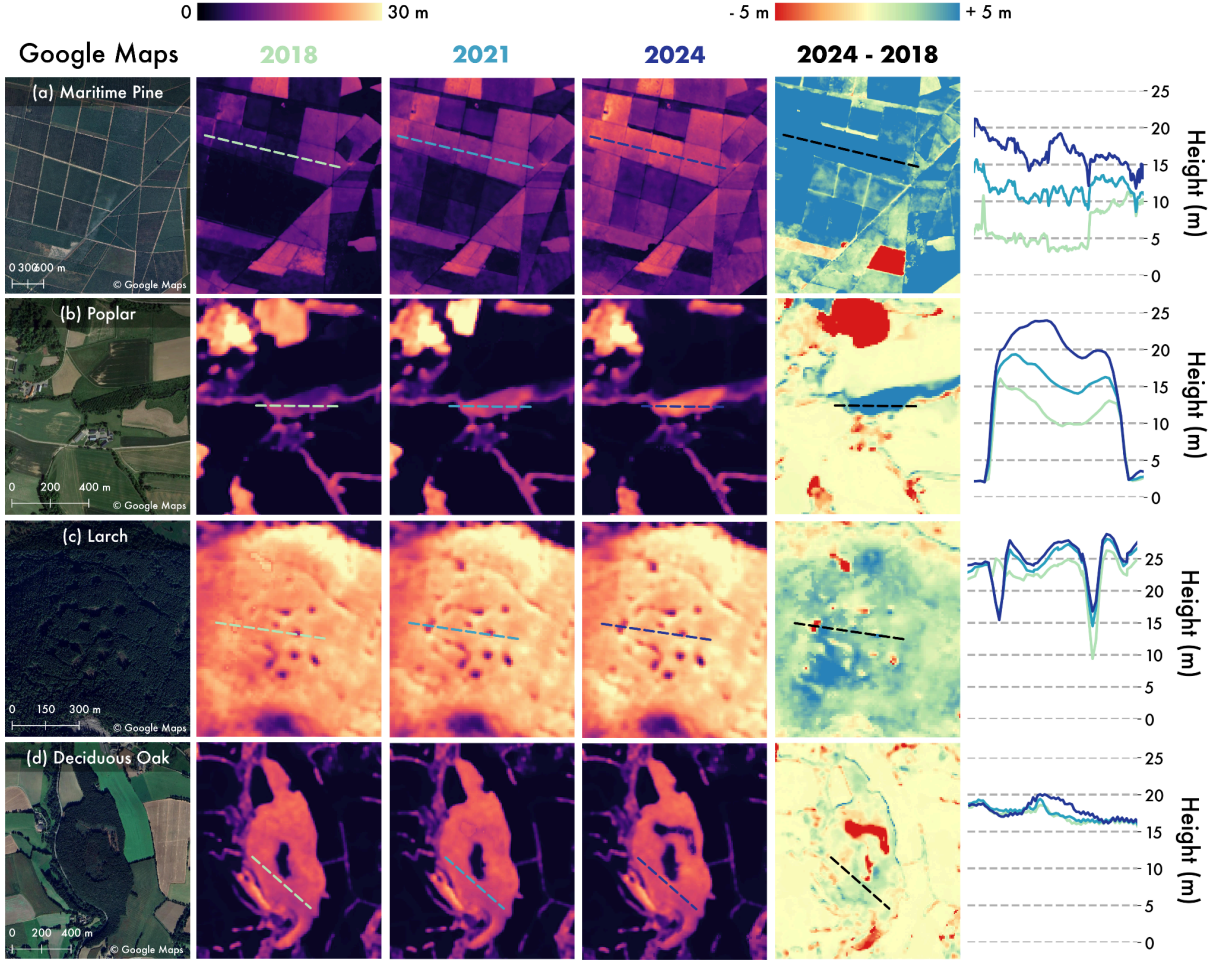

Fig 6: Forest height in 2018, 2021, and 2024; Google Maps © images; and the difference between 2024 and 2018. The last column shows the height profiles drawn on the maps. The four lines correspond to different tree species: (a) Maritime pines (44.54 N, -1.03 E) (b) Poplar plantation (49.72 N, 3.86 E) (c) Larches (44,98 N, 6.47 E) (d) Deciduous oaks (48.30 N, -3.61 E). Adapted from (Schwartz, 2023)

Accurately mapping forest biomass is one of many tasks that can be addressed using Earth Observation (EO) data combined with deep learning algorithms. The rise of very high-resolution satellite imagery—such as Maxar satellites that go up to 15 cm resolution, SPOT satellites (1.5 m resolution), and the Planet Labs constellation, which provides 3.5 m resolution images of the entire globe daily—offers new opportunities to analyze forest structure changes at the tree level. Some of these private data are made available for research purposes, such as the annual SPOT mosaics in France that have been used to produce the *open-canopy* maps from (Fogel et al., 2024). These annual 1.5 m resolution maps have proven a high accuracy in detecting individual tree removal in France's forests. They could significantly enhance our understanding of forest disturbances and improve the existing attribution of forest losses (Fogel et al., 2024; Viana-Soto and Senf, 2024). By analyzing disturbance patterns, they could help differentiate between natural forest losses—such as windthrows, wildfires, and pest outbreaks—and human-driven factors like selective logging, clear-cuts, and salvage logging.

Another key application is the creation of tree species maps (Beloiu et al., 2023), which could refine height-biomass allometries and improve our understanding of forest diversity within a region. By analyzing temporal information from satellite time series, recent studies have successfully identified tree phenology and linked it to specific species, as demonstrated in Belgium by (Bolyn et al., 2022). With large labeled datasets like the Pure Forest dataset (Gaydon and Roche, 2024), future research can fully exploit remote sensing data using advanced deep learning models such as Vision Transformers (Dosovitskiy et al., 2020). These high-precision tree species maps would significantly enhance forest management, monitoring, and carbon stock estimation.

While advances in remote sensing and machine learning are transforming forest monitoring, ground-truthing remains essential for accurate calibration and assessment. For example, growth measurements, crucial for estimating carbon stocks, are only possible in open canopies, requiring in-situ circumference tracking as forests mature (Wernick et al., 2021). Another promising approach to upscaling mortality prediction is linking ICP (International Co-operative Programme on Assessment and Monitoring of Air Pollution Effects on Forests) crown defoliation data. A recent study using defoliation data from Scots pines in Switzerland shows that ground-assessed defoliation rates serve as early-warning signals for mortality (Hunziker et al., 2022). These insights could be expanded using remote sensing and machine learning, integrating defoliation monitoring with high-resolution satellite data like Sentinel-2 and GEDI LiDAR. Similarly, Sentinel-2 can be used for early detection of bark beetle outbreaks but still need systematic ground-truthing (REF) ("Early detection of bark beetle infestation in Norway spruce forests of Central Europe using Sentinel-2," 2021). Additionally, citizen science can play a critical role, as initiatives such as *Santé des Forêts* already mobilize volunteers to report pest outbreaks, creating a valuable database for linking mortality to biotic disturbances and climatic stress. Public records, analyzed with large-scale language models, could further help reconstruct past disturbance events (REF). However, while accurate geolocation of mortality events is essential for attribution (e.g., the French NFI provides coordinates only within a 700m radius of the true location, a prevalent practice across world NFIs, (Schadauer et al., 2024) , with persisting debates about making such data openly available (Gessler et al., 2024). For those NFI placed under the authority of public statistics like in France, coordinate access management will be managed according to the law on statistical secrecy. Another - perhaps more crucial - issue is that NFI protocols based on nested plots (not all the trees being measured) are not primarily intended for remote-sensing calibration, in contrast with experimental forest plots (e. g. Seynave et al. 2018, AFS) (Seynave et al., 2018).

However, the statistical robustness of the systematic sampling design used in the French National Forest Inventory, which ensures unbiased estimates at large scales, makes it a cornerstone of forest observation. On the one hand, such a tool can be used for calibration and validation of RS products, but on the other, it can also benefit from these to both satisfy increased precision and mapping requirements (Kangas et al., 2018). Indeed, the classical NFI can be supplemented by AI-powered and all remote sensing-based products aforementioned, through the concept of multi-Source forest inventory, an innovation rewarded by the Marcus Wallenberg Prize in 1997 (Tomppo et al. 2008, book in SPringer) and matter of ongoing developments in France (Vega et al., 2021) (Sagar et al., 2022). A strong advantage here is that RS products play an auxiliary role in a design-based statistical approach, and do not need to be unbiased, but informative. Therefore, a virtuous integration can be figured out between NFI and RS product developments, the former being used for calibrating the latter in a high-frequency and high-resolution perspective, while the latter will usefully increase the precision and frequency of the former (Bontemps et al. 2022). Beyond the model-assisted paradigm, the capability of AI to improve the correlation between remote sensing and NFI will reduce the risk of bias associated with model specifications and enhance the capabilities of model-based small area estimation for the estimating of small scale phenomena such as local disturbances. Given the demonstrated complementarity of these AI-powered and remote sensing-based models (Besic et al., 2025; Sagar et al., 2022), we can confidently state that holistic monitoring approaches - combining satellite observations, AI, and structured field data - will be essential for improving mortality predictions and understanding forest carbon dynamics and forest monitoring in general (Bontemps et al. 2022).

**Concluding remarks and recommendations**

We have shown that the forests in France have experienced an alarming decline of their CO2 sink which went from 49.3 MtCO2 yr$^{-1}$ in 1990 to 37.8 Mton CO2 yr$^{-1}$ in 2022. This decrease took place mainly from 2013 to 2017 and seems to be associated with an abrupt rise of both tree natural mortality related to climate change drivers and to a parallel rise of harvests. This harvest increase could reflect salvaged trees after natural dieback. In recent years, two northern regions have become net CO2 emitters to the atmosphere and other regions have CO2 fluxes that decreased and are now close to zero. In contrast, southern regions remained stable CO2 sinks, be they intensively managed like in Les Landes or lightly managed mediterranean forests. These findings make the country's carbon neutrality target out of reach for the period, as it was based in the initial SNBC (2020) based on the Land Use, Land Use Change and Forestry data from 2018 which set as a target a sustained a sink of 41 MtCO2 yr-1 in the forest sector in 2015, then -39 MtCO2 yr-1 in the second budget 2019-23. This objective became unachievable in light of the alarming decrease of the CO2 sink, and has been revised to -9 MtCO2 yr-1 in 2024 (Fig. 1b). The new objective may still be challenging to reach in light of continuing greenhouse gas emissions in the agriculture sub-sector and rising tree mortality in the coming years.

Our findings show that tree mortality has become a dominant driver of the declining carbon sink in France, with significant regional differences and species-specific vulnerabilities. The acceleration of mortality since 2013, independent of harvesting trends, suggests that climate-induced stressors—especially droughts and biotic disturbances—are reshaping forest dynamics. Given the delayed detection of mortality in national forest inventories, complementing field-based observations with high-resolution remote sensing and citizen science initiatives could provide earlier warning signals and improve attribution of mortality drivers. Strengthening long-term monitoring networks and integrating predictive modeling will be critical to managing forests more effectively in a changing climate.

The national inventory is a unique resource to monitor long term carbon trends in forests and forms one of the bases for the national reporting of France to the UNFCCC. However, the current sampling scheme does not allow for year on year monitoring of changes, and cannot attribute the location and severity of carbon losses occurring through disturbances as soon as these are small-scale. In this respect, satellite imagery provides excellent coverage down to tree level but requires ground data to provide robust maps of carbon accounting. We make the following recommendations : 1) make exact locations of the national inventory samples available to the research community, in the respect of public legislation on private data protection, so that these precious data can be combined with satellite imagery and analysis to improve the monitoring of forests in France, 2) provide publicly available geospatial information on forest clearcuts, and natural disturbances such as insects, drought mortality events based on the best available science, 3) measure deadwood changes as an overlooked carbon stock in the French forests, 4) develop artificial intelligence and data integration methods to monitor gains of forest carbon from satellite and airborne imagery.

Also, the present aim of tracking changes in the forest carbon sink and its components over a long-term period (1990-2022) has not been frequently addressed to date, and it highlighted clear limitations associated to the different tools contributing to this monitoring, should it be accurate and homogeneous over time. These limitations call for some minimum caution in the interpretations of the results, and also come along with recommendations to the national community:

(i) **Forest harvest-reconstitutions over the long term** - before 2010, no accurate field monitoring of forest removals was implemented, making the NFI data non-operative in this respect, and requiring external and indirect quantification approaches (including the *Enquête Annuelle de Branche*, for wood sawn in the industries) routinely using for forest carbon reporting. With forest growing stock being however estimated with a high precision (around 1% at country scale, and at a much higher precision in former decennial inventories), it is suggested that forest harvests be retrospectively reconstituted by equating the annual difference of carbon stocks (net carbon sink) to the sum of forest fluxes. The approach has been successfully tested over ancient departmental inventories (Denardou 2019), and may be generalized to all available inventories.

(ii) **Delivering long time series of forest attributes** - Ancient inventories were based on asynchronous departmental surveys, which forms a systematic limit for a rapid production of forest time series at different scales at which statistical precision is sufficient (GRECO/SER ecological classifications, or NUTS-3/NUTS-2 levels of EU administrative units). Using interpolation and aggregations, time series of forest state variables (area, total stock and biomass) and fluxes (growth, harvest, mortality) should also be produced for these different classifications. The linearity of these reconstitutions make the delivery of associated confidence intervals prone to limited developments.

**(iii) Updating the wood density component of the aerial forest carbon sink** - Highly diversified forests of the French territory pose a sharp limit on the accuracy of wood density quantification across tree species (>150 found at the censusable stage in France), for which the Carbofor project remains a cornerstone, some density records however remaining doubtful. Yet, an unprecedented effort relying on a collaboration between IGN and INRAE has been consented since 2015, with a view to establish a systematic wood density record for the French forests, based on the sampling design of the national forest inventory. These open data (Cuny et al., 2025)should be used whenever aerial tree carbon stocks/sinks are requested, and may lead to a positive reevaluation of the forest carbon.0

**(iv) Tracking for total natural tree mortality in forests** - Whereas mortality was the matter of restricted concern up to one decade, its abrupt increase after severe drought or pest-related disturbances has drawn attention onto this flux (e g. Taccoen et al. 2019) (Taccoen et al., 2019). In spite of major improvements introduced in the French NFI with the semi-permanent design, 5 years can remain limiting for mortality assessment after massive events. This is at a risk of both underestimating natural mortality, and overestimating planned harvests. Complementary perspectives rely on both progress in forest inventory design (e. g. spatially systematic panelling of plot revisits across 5 years has no cost difference), and developments in remote sensing technologies, from which one may expect bias-correction options in view of the higher frequency of these products. For these reasons, it is urged to amplify the scientific dialog between remote sensing and forest inventory science.